\documentclass[acmsmall]{acmart}

\usepackage{lineno}

\newcommand{\Z}{\mathbb{Z}}

\newcommand{\R}{\mathbb{R}}

\newcommand{\Vc}[1]{{\boldsymbol #1}}

\newcommand{\doublefigure}[5]{{\begin{figure}[htb]%
\centering %
\includegraphics[width=#1\linewidth]{#2}
\includegraphics[width=#1\linewidth]{#3}
\caption{#4}
\label{#5}%
\end{figure}}}

\newcommand{\singlefigure}[4]{{\begin{figure}[htb] %
\centering %
\includegraphics[width=#1\linewidth]{#2}%
\caption{#3}%
\label{#4}%
\end{figure}}}

\AtBeginDocument{%
  \providecommand\BibTeX{{%
    \normalfont B\kern-0.5em{\scshape i\kern-0.25em b}\kern-0.8em\TeX}}}

\setcopyright{acmcopyright}
\copyrightyear{2023}
\acmYear{2023}
\acmDOI{}

\acmJournal{JACM}
\acmVolume{--}
\acmNumber{--}
\acmArticle{---}
\acmMonth{3}

\begin{document}

\title{Strassen's Matrix Multiplication Algorithm Is Still Faster}

\author{Paolo D'Alberto}
\email{}
\affiliation{%
  \institution{AMD}
  \streetaddress{2100 Logic Dr}
  \city{San Jose}
  \state{California}
  \postcode{95124}
}

\renewcommand{\shortauthors}{D'Alberto et al.}

\begin{abstract}
  Recently, reinforcement algorithms discovered new algorithms that
  really jump-started a wave of excitements and a flourishing of
  publications. However, there is little on implementations,
  applications, and, especially, no absolute performance and, we show
  here they are not here to replace Strassen's original fast matrix
  multiplication yet.  We present {\em Matrix Flow}, this is a simple
  Python project for the automatic formulation, design,
  implementation, code generation, and execution of fast matrix
  multiplication algorithms for CPUs, using BLAS interface GPUs, and
  in the future other accelerators. We shall not play with module-2
  ($\Z_2$) algorithms and, for simplicity, we present only square
  double-precision matrices. By means of factorizing the operand
  matrices we can express many algorithms and prove them
  correct. These algorithms are represented by Data Flows and matrix
  data partitions: a Directed Acyclic Graph. We show that Strassen's
  original algorithm is still the top choice even for modern GPUs. We
  also address error analysis in double precision, because integer
  computations are correct, always.
\end{abstract}

\maketitle

\section{Introduction} 
\label{sec:introduction}
In the literature, there are quite a few algorithms for {\em fast}
matrix multiplication already. For example, we used to have a few
pages of references showing the rich literature. Now institutions and
companies are actually stepping in and providing a complete
list. Although the list of algorithms is long, the list of
implementations for old and new architectures is limited in absolute
terms. Here and always, we are interested in the computation with
matrices, not a single elements or scalar arithmetic, and we compare
the trade between matrix multiplications for {\em pre} and {\em post}
matrix additions.  This is important when we translate fast algorithms
into practical implementations. The temporary spaces required is also
an interesting aspect of the algorithm performance.

The researchers at DeepMind presented a new methodology to find new
and faster matrix multiplications \cite{PMID:36198780}. They provide
the algorithms in matrix forms and, easily, we can generate, and
execute element-wise algorithms.  There are new algorithms and the
authors show the relative advantage performance with respect to the
classic matrix multiplication for GPUs and TPUs (using XLA).

Here, we show that Strassen's algorithm \cite{STRASSEN1969} is still
king of the hill.  We explore a comparison across several algorithms
using the standard arithmetic: double precision matrices and no $\Z_2$
modular arithmetic because it is the default in Python. Half, single
and double precision, single and double complex, and also integer
8-bit can be deployed when the BLAS interface defines it for an
architecture. In this scenario, we can determine a priori the relative
advantages of all algorithms (i.e., 7\% improvement per
recursion). More interesting, we can show that {\em deeper} algorithms
such as the one recently discovered do not have all the advantages
when other constraints are present.

In the following, we present a Python project: {\em Matrix
  Flow}. Here, you can take any of the algorithms for square matrices,
we use the DeepMind's algorithms as repository, and create complex
implementations you can run in CPU, GPU and in principle extend to any
accelerator using the appropriate interface: we use BLAS GEMM $*$ and
GEMA $+$, these are the minimum operations for the computation in a
non commutative ring. We can use also Bini's format for the
representation of the algorithms, which is our first and preferred
\cite{Paolo2007}.

We have a few goals in mind for this work. First, we want to make
available algorithms and Python implementations: both proven
correct. Second, we want to provide tools for the implementation of
these algorithms for highly efficient systems. Python is flexible,
with such a nice abstraction level that writing code for matrix
algorithms is a breeze. Efficiency and performance is not always
Python forte, however writing interfaces for C/C++ is relatively
simple. Interfaces are not a very good solution either and they are
more a nice intermediary solution. Think of accelerators where
interfaces have to change the shapes, layouts of the operands, and
send them through a slow connections. So we provide tools to build
entire algorithms in C/C++ for CPU and GPUs and still prove them
correct and quantify rounding errors. Last, we provide tools for the
validation of algorithms.

So this is a paper and a software project introduction. Here, we would
like to share a few lessons learned that are not trivial, sometimes
original, and important for other applications where compiler tools
are designed to optimize artificial intelligence applications.

\section{Matrices and Partitions.}
\label{sec:matrices}
We start by describing the non commutative ring: matrices, operations,
and matrix partitions. This will help us understand and describe fast
matrix multiplications constructively.

In this work we work mostly with matrices and scalar, but vector are
easily added. A matrix $\Vc{A}$, this is a square matrix of size
$\R^{n\times n}$. Every element of the matrix is identified by a row
$i$ and a column $j$ as $\Vc{A}_{i,j}$. A scalar $\alpha$ is number in
$\R$.
\begin{itemize}
  \item $\Vc{B} = \alpha\Vc{A} = \Vc{A}\alpha$ so that $\Vc{B}_{i,j} =
    \alpha\Vc{A}_{i,j}$ and these are (scalar) multiplications in
    $\R$.
  \item $\Vc{B} = \alpha\Vc{B} + \beta\Vc{C}= \beta\Vc{C} +
    \alpha\Vc{B}$ is matrix addition and $\Vc{B}_{i,j} =
    \alpha\Vc{B}_{i,j} + \beta\Vc{C}_{i,j}$ and it is commutative.
  \item $\Vc{B} = \alpha(\Vc{B} + \Vc{C}) = \alpha\Vc{B} +
    \alpha\Vc{C}$. 
  \item $\Vc{C}= \Vc{A}\Vc{B}$ is matrix multiplication and $\Vc{C}_{i,j}=$ 
    $\sum_{k=1}^n \Vc{A}_{i,k}\Vc{B}_{k,j}$, it is not commutative.
  \item ( $\Vc{y}= \alpha\Vc{A}\Vc{x}$ is matrix by vector
    multiplication and $\Vc{y}_{i}=\alpha \sum_{k=1}^n
    \Vc{A}_{i,k}\Vc{b}_{k}$, you will find this in the software
    package but it is not necessary).
\end{itemize}

Great, we have the operands and the operations. Let us introduce the
last component for the description of our algorithms: Partitions.  The
double subscript for the definition of elements is for exposition
purpose only. Here, we introduce the single subscript that in the
original literature is used a lot (if not exclusively). Consider a
matrix $\Vc{A} \in \R^{n\times n}$, we can consider the matrix as a
composition of $\Vc{A}_i \in \R^{\frac{n}{2}\times \frac{n}{2}}$
sub-matrices.
\begin{equation} 
  \Vc{A} =
  \begin{pmatrix}
    \Vc{A}_0 & \Vc{A}_1 \\
    \Vc{A}_2 & \Vc{A}_3 \\
  \end{pmatrix} = \{ \Vc{A}_0 , \Vc{A}_1, \Vc{A}_2 , \Vc{A}_3  \}
\end{equation}
The property is that $\Vc{A}_i$ are non overlapping matrices and this
is easily generalized to any factor of $n$ for example $p$ so that $n
= k*p$. For $p=3$ and $n=3k$ and $\Vc{A}_i \in \R^{\frac{n}{3}\times
  \frac{n}{3}}$ and $0\leq i\leq p^2 -1$:
\begin{equation} 
  \Vc{A} =
  \begin{pmatrix}
    \Vc{A}_0 & \Vc{A}_1 & \Vc{A}_2 \\
    \Vc{A}_3 & \Vc{A}_4 & \Vc{A}_{5} \\
    \Vc{A}_6 & \Vc{A}_7 & \Vc{A}_8 \\
  \end{pmatrix}\\
  = \{ \Vc{A}_0 , \Vc{A}_1, \Vc{A}_2 , \Vc{A}_3 ,\Vc{A}_4 , \Vc{A}_5, \Vc{A}_6 , \Vc{A}_7, \Vc{A}_8   \}
\end{equation}
Note, when we have a partition $\{ \Vc{A}_i \}$ we know the matrix
$\Vc{A}$ completely.  A partition recursively applied reduces to a
list of scalar elements. Let us represent the simplest matrix
multiplication using $2x2$ partition:

\begin{equation}
  \label{basic}
  \begin{pmatrix}
    \Vc{C}_0 & \Vc{C}_1 \\
    \Vc{C}_2 & \Vc{C}_3 \\
  \end{pmatrix} =
  \begin{pmatrix}
    \Vc{A}_0 & \Vc{A}_1 \\
    \Vc{A}_2 & \Vc{A}_3 \\
  \end{pmatrix} *
  \begin{pmatrix}
    \Vc{B}_0 & \Vc{B}_1 \\
    \Vc{B}_2 & \Vc{B}_3 \\
  \end{pmatrix} 
\end{equation}
Using the factor $p=2$, we have in general
\begin{equation}
  \label{eq:recursion}
  \begin{pmatrix} 
    \Vc{C}_0 = \sum_{k=0}^{1}\Vc{A}_k    \Vc{B}_{2k} &
    \Vc{C}_1 = \sum_{k=0}^{1}\Vc{A}_k    \Vc{B}_{2k+1} \\
    \Vc{C}_2 = \sum_{k=0}^{1}\Vc{A}_{2+k} \Vc{B}_{2k} &
    \Vc{C}_3 = \sum_{k=0}^{1}\Vc{A}_{2+k} \Vc{B}_{2k+1} \\
  \end{pmatrix}
\end{equation}
Equation \ref{eq:recursion}, as matrix computation, represent
naturally a recursive computation. Take the $\Vc{T}_0=\Vc{A}_0*\Vc{B}_0$, we
can compute it directly or partition the operands further not necessarily in the same way as in Equation \ref{basic}.  

A Strassen's algorithm has the following format:
\begin{equation}
  \label{strassen}
C_i = \sum_{j=0}^6 c_{i,j}P_j =\sum_{j=0}^6 c_{i,j}T_j*S_j  
=\sum_{j=0}^6 c_{i,j}\Big[ T_j = \big(\sum_{k=0}^3
  a_{k,j}A_k\big)\Big]* \Big[S_j = \big(\sum_{\ell=0}^3 b_{\ell,j}B_\ell\big) \Big]
\end{equation}
There are seven products, each product can be done recursively, but
first we must combine submatrices of the operands accordingly to a
specific set of coefficients. The coefficients $c_{i,j},a_{k,j},$ and
$b_{\ell,j}$ belong to three sparse matrices. Let us take the
Strassen's algorithm but represented as DeepMind would use it for the
computation of Equation \ref{strassen}. We choose to use single
subscripts for $C_i, A_j, B_k$, because the notation is simpler (i.e.,
fewer indices), and because we implicitly use a row major layout of
the sub-matrices making the single index completely determined.

\begin{equation} 
  \Vc{a}, \Vc{b}, \Vc{c}^t = f[`2,2,2`] \\
\end{equation}
Where the matrices $\Vc{a}, \Vc{b}$, and $\Vc{c}$ are integer matrices
with coefficients [-1,0,1], but they do not need to be integers as
long the matrix stays sparse, described as follow plus some useful
notations (the first column and the last row are for notations purpose
only to connect to Equation \ref{strassen}):

\begin{equation}
  \Vc{a}= 
  \begin{pmatrix}
    A_0 & 0   & 1  & 1 &  0 &  1&  1&  0 \\
    A_1 & 0   & 0  &-1 &  1 &  0&  0&  0 \\ 
    A_2 & 1   & 1  & 1 &  0 &  1&  0&  0 \\
    A_3 &-1   &-1  &-1 &  0 &  0&  0&  1 \\ \hline
        & T_0 &T_1 &T_2 &T_3&T_4 &T_5&T_6 \\
  \end{pmatrix},
  \Vc{b}= 
  \begin{pmatrix}
    B_0 &0   & 0 & 0 & 0 & 1 & 1 & 0 \\
    B_1 &1   & 1 & 0 & 0 & 1 & 0 & 1 \\
    B_2 &0   & 1 & 1 & 1 & 1 & 0 & 0 \\
    B_3 &0   & 1 & 1 & 0 & 1 & 0 & 1 \\ \hline
        &S_0 &S_1&S_2&S_3 &S_4&S_5&S_6 \\
  \end{pmatrix},
\end{equation}
and

\begin{equation}
  \Vc{c}^t =
  \begin{pmatrix}
    C_0 &  0&  0&  0&  1&  0&  1&  0 \\
    C_2 &  0& -1&  0&  0&  1& -1& -1 \\
    C_1 & -1&  1& -1& -1&  0&  0&  0 \\
    C_3 &  1&  0&  0&  0&  0&  0&  1 \\ \hline
        &P_0&P_1&P_2&P_3 &P_4&P_5&P_6 \\
  \end{pmatrix} 
\end{equation}

Take matrix $\Vc{c}$ and consider the first row ($C_0$).
\begin{equation}
  \Vc{C}_0 = P_3+P_5 = (T_3*S_3) + (T_5*S_5) = (1A_1)*(1B_3) + (1A_0)*(1B_0)
\end{equation}
and for better effect the last row ($C_3$).
\begin{equation}
  \Vc{C}_3 = P_0+P_6 = (T_0*S_0) + (T_6*S_6) = (A_2-A_3)*(B_1) + (A_3)*(B_1+B_3)
\end{equation}
If we take the matrix $\Vc{c}^t$, the columns are the products of the
algorithms, in this case 7. The row explains how to sum these product
so that to achieve the correct result. The matrix $\Vc{a}$, each
column $i$ represents the left operand of product $P_i$. The matrix
$\Vc{b}$ represent the right operand. Notice, that with the proper
matrices we could achieve Equation \ref{eq:recursion}

We shall use only square matrices for simplicity. Thus the key for the
full determination of the Strassen algorithm is 2. This is the common
factor we use to split both sides (i.e., row and column) for all
matrices. If the split factor determines the algorithm with minimum
number of products, so we are going to {\em play} with 2 as 7
products, 3 as 23 products, 4 as 2x2 and 49 products, 6 = 2x3 and 3x2
as 7*23 products, 9 = 3x3 as 23*23 products, 12=2x2x3, 2x3x2, 3x2x2 as
7*7*23 products.

Intuitively and if we are using the recursive idea as in Equation
\ref{strassen}, take the algorithm with factor 6. We can take fist the
factor 2, and each of the 7 products, recursively use an algorithm
with factor 3 and 23 products. A pure recursive algorithm would
compute the operands $T_i$ and $S_i$ first, recursively solve the
result $P_i$ and distribute it. The sequence 2 and 3 specifies an
algorithm that is different from the sequence 3 and 2.  The space
requirement is different: literally if we take a 6x6 matrix, if we
split the matrices by 2, the temporary space to hold $T_i$ and $S_i$
are two matrices of size 3x3. The recursive call will use temporary
space of size 1x1 (scalar computation). Otherwise, if we split by 3,
we need a temporary space of size 2x2. The computation order is
different and, even in double precision, there is a different error
distribution.

Interestingly, the original problem (6x6x6), which is the complete
reference in \cite{PMID:36198780} splits the problem $(M,N,K)$ into
$(\frac{M}{6},\frac{N}{6}\frac{K}{6})$ basic products can be done
using non square ones: (2x3x3) and (3x2x3). Thus, we can solve
directly using a factor 6 and square matrices but with a proper
factorization of $\Vc{a},\Vc{b}$, and $\Vc{c}$ and the order is
important.

In the project there is a complete proof and a few implementations: In
practice, if we use a factor 3 and we have $\Vc{a}_3, \Vc{b}_3,
\Vc{c}_3$ and we have a factor 2 with $\Vc{a}_2, \Vc{b}_2, \Vc{c}_2$,
then the algorithm with factor 6 is completely defined succinctly as
$\Vc{a}_3 \otimes \Vc{a}_2, \Vc{b}_3 \otimes \Vc{b}_2, \Vc{c}_3
\otimes \Vc{c}_2$. Where $\otimes$ is the Kronecker product (used also
in \cite{PMID:36198780}). This factorization and algorithm generation
is really neat: If we have a problem size that can be factorized
exactly with the set of fast algorithms, then it requires just
constant temporary space (last factor squared that can be just 1). We
shall show application of this idea in the experimental results.
\newpage 
\section{Matrix Flow}
\label{sec:matrixflow}

Take a matrix operation such as $\Vc{C} = \alpha \Vc{A}\Vc{B}$, and
describe this computation by means of the matrix partitions by
factors: recursively if you like by a factor of 2
\begin{equation}
  \label{basic_2}
  \begin{pmatrix}
    \Vc{C}_0 & \Vc{C}_1 \\
    \Vc{C}_2 & \Vc{C}_3 \\
  \end{pmatrix} = \alpha
  \begin{pmatrix}
    \Vc{A}_0 & \Vc{A}_1 \\
    \Vc{A}_2 & \Vc{A}_3 \\
  \end{pmatrix} *
  \begin{pmatrix}
    \Vc{B}_0 & \Vc{B}_1 \\
    \Vc{B}_2 & \Vc{B}_3 \\
  \end{pmatrix} 
\end{equation}
If we use the definition of matrix multiplication to perform the
computation, we can build a sequence of instructions as follows:
{\small \begin{verbatim}
    ## Declarations AD single index partition of A, BD of B, CD of C 
    decls = [[alphai],    AD ,        BD,          CD ]     
    ###
    ## Computation as a sequence of assignment statements and binary operations.  
    ###
    V = []  ## operation list  
    for i in range(Row=2):
        for j in range(Col=2):
            T = Operation("p0", "*", AD[i*Col] ,BD[j])
            for k in range(1,K=2):
                T1 = Operation("p%d"%k, "*", AD[i*Col+k],BD[k*Col+j])
                T  = Operation('c','+',T,T1)
            R = Operation("s0", '<<',CD[i*Col+j],T)
            V.append(R)
\end{verbatim}}
{\em Operation} is simply an abstraction of the matrix operations $*$
and $+$, we have only binary computations plus a constant
multiplication. Such a format is standard in BLAS libraries. In
practice, the declarations and the instruction sequence (i.e., V)
define the computation completely. In matrix flow this is a {\em
  Graph}. A Graph is an extension of a {\em function} where there can
be multiple inputs and outputs.  Think about the factorization of a
matrix into two components $\Vc{L},\Vc{U} = lu(\Vc{A})$. A function is
an extension of a matrix {\em operation}: one output and two operands.

The idea behind a graph is simple: we have clearly spelled out inputs
and ouputs, we can build a clear declaration of the matrix partitions
and constants, we have a list of assignment statements.  The
computation is summarized by a directed acyclic graph and the schedule
by the sequence of instructions based on binary operators represented
as an abstract syntax tree.

In the last ten years, the compiler technology for data flows
computations and DAGs has had great incentives because it is the
foundations of artificial intelligence applications for CPU but
especially for custom accelerators (among them GPUs).

The main features of a graph in combination of using Python at the
time of this writing are the following.
\begin{enumerate} 
\item The self description of the graph is a readable code. If we
  create a graph using Equation \ref{basic_2} , the print of the graph
  is literally {\small
\begin{verbatim}
  (Pdb) print(G1)
  CDP[0] << (ADP[0]) * (BDP[0]) + (ADP[1]) * (BDP[2])
  CDP[1] << (ADP[0]) * (BDP[1]) + (ADP[1]) * (BDP[3])
  CDP[2] << (ADP[2]) * (BDP[0]) + (ADP[3]) * (BDP[2])
  CDP[3] << (ADP[2]) * (BDP[1]) + (ADP[3]) * (BDP[3])
\end{verbatim}
}
The symbol \verb2<<2 is the assignment operation and it is overwritten
instead of the Python shift operation (better than overwriting the
assignment itself).

\item The self description is valid Python code and it can be compiled
  and executed. So this is completely similar to LLVM where the graph
  can be explicitly encoded, the code is human readable, and it can be
  modified (the graph), but we can {\em execute} the code directly.

\item We can compute the graph by executing the list of operation in
  sequence.  Can you see the list $V$ in the snippet? This is a list
  of operations we can evaluate sequentially. So the graph is a
  computational data structure. The list is one valid schedule and we
  can change it.

\end{enumerate}

In summary, the graph is a self contained structure that can be used
to execute the algorithm, to compile it, and thus validate easily. Our
goal is to describe how we use this to validate correctness and
relative performance of any fast algorithms and to show that by
customizing the description of the graphs as code, we can generate
native codes for CPU and GPU that can be easily compiled and validated
in the same environment.  Thus can be validated theoretically and
practically. We want to focus on how we can build efficient codes
especially for GPUs.

\subsection{CPU and Python performance and validation}

We have a systematic way to retrieve fast algorithms by factors, we
can create computational graphs, for a specific problem size, and we
can execute them. The first thing it is to get a feel of the
performance, like in Figure \ref{fig:perf1}.

\doublefigure{0.50}{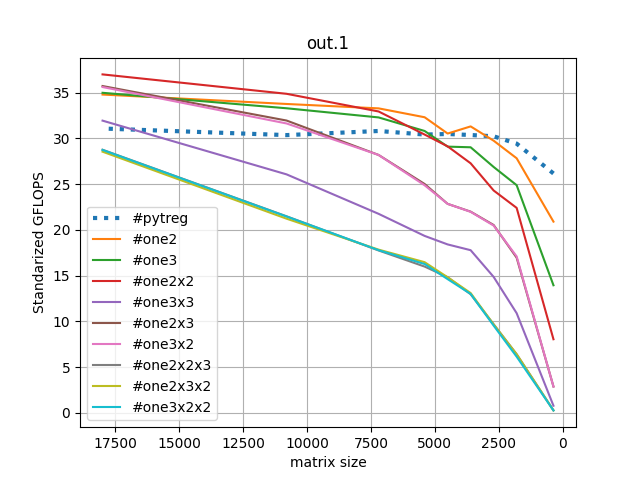}{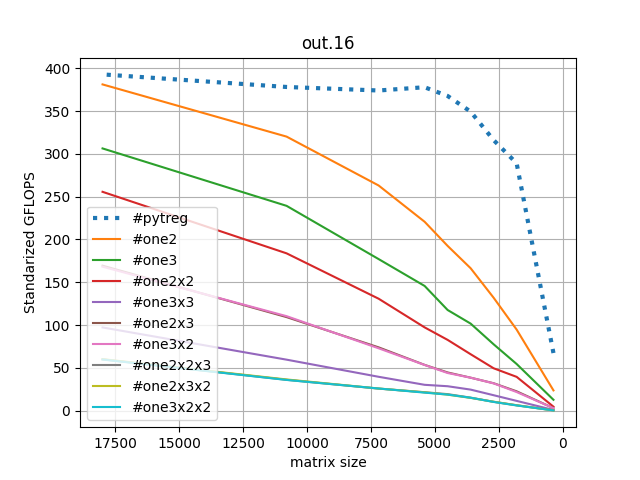}{Performance with 1 CORE
  and 16 cores from GEMM}{fig:perf1}

There is a lot to unpack. There are two plots, the matrix sizes go
from the largest to the smallest so we can highlight the maximum speed
up as the first spot in the graph.  This is a Thread-ripper CPU with
16 cores. We can afford to give one core to the base line GEMM (and
for all fast algorithms) and we can give them 16. The first plot show
the performance when we use 1 core.  The Base line is steady at 30
GFLOPS. It is clear that Strassen 2 and 2x2 (respectively,
\verb3#one23 and \verb3#one2x23) is faster for any problem size 2500
and above. All algorithms with three or more factors are behind and
they never break even but the trajectory is promising ( \verb4#one3x2x24).

The situation is different as soon as the GEMM gets more cores: In the
second plot no algorithm is faster than the base line. The reason is
the base line does not have a sharp increasing and, then, steady
performance; that is, the performance knee between sizes 0 and
5000. In practice, the additions are constantly bad performing and the
multiplications we save are small and inefficient, we are better off
solving a large problem using the standard GEMM. In practice, all fast
algorithms seem having slopes for which there will be no intersection
or crossing for no foreseeable size.

Of course, the 16 core example is skewed towards the single call GEMM
algorithm, which is never slowed down by the several function calls
for each operations.

\subsection{GPUs}
Matrix Flow can use the same methodology to create C++ code using
rocBLAS interface. In practice, we create rocBLAS calls into a single
C++ package that we can compile and import in Python. Again, the
interface and module idea is not efficient, it is not meant to replace
any of the current implementations in C/C++/Fortran. It is just so
much easier to compare and validate the correctness of the computation
and report the numerical stability of the code.

\singlefigure{0.60}{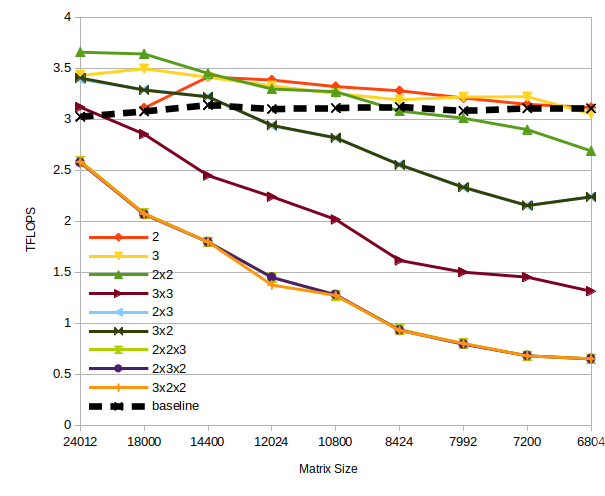}{Performance with VII Radeon 3.5TFLOPS
  DGEMM}{fig:perf2}

The GPU we use here is a Vega20 known as VII Radeon. The performance
is for double precision and the peak performance is 3.5TFLOPS (tera =
$10^{12}$ floating point per second). This GPU has a 16GB of HBM2 main
memory with 1TB/s bandwidth.  In Figure \ref{fig:perf2}, we show the
performance again from the largest to the smallest: so we can
appreciate better the maximum speed up for the fast algorithms. We
report {\em \bf kernel} wall clock time that is the computation of the
sequence of GEMM and GEMA without any CPU device compaction and
transmission. In general, the communication between CPU and GPU takes
so much more than the computation itself.

The good news is that the GPU performance is identical to the 1CORE
plots (instead of the 16). Once again three factors algorithm are
behind but the trajectory is pretty good they are close to break even
(we did not investigate if they do because the largest problem we can
execute is about 26000x26000 without leaving any space to the GEMM
kernel). The 2-factor algorithms go above the baseline and Strassen 2
and 2x2 are safely in the lead.

Two caveats: the sizes we test are designed so that all algorithms
have the proper factors but this does not mean that are suitable for
the GEMM kernel. For the last problem size, 24012, the problem size is
about 13GB (which was measured at about 80\% of the VRAM
available). Due to the temporary space, Strassen with factor 2, does
not fit into the device (requiring about extra 3GB).

\section{Conclusion}
As short conclusion, in $\R$ and with {\em optimal} implementation of
GEMM and GEMA, deeper algorithms like the one recently found by
DeepMind do not have the edge against the classic Strassen's
algorithm. If you think about it, Strassen is still pretty good.


\bibliographystyle{IEEETran} \bibliography{ref}

\end{document}